\documentclass[12pt,preprint]{aastex}
\usepackage{graphicx}
\usepackage{epstopdf} 
\usepackage{amsmath}

\begin{document}

\title{Adaptive Optics Images II: 12 Kepler Objects of Interest and 15 Confirmed Transiting Planets\altaffilmark{1}}

\author{E. R. Adams\altaffilmark{2}, A. K. Dupree\altaffilmark{2},  C. Kulesa\altaffilmark{3}, D. McCarthy\altaffilmark{3}}

\altaffiltext{1}{Based on observations obtained at the MMT Observatory, a joint facility of the Smithsonian Institution and the University of Arizona.}
\altaffiltext{2}{Harvard-Smithsonian Center for Astrophysics, 60 Garden St., Cambridge, MA, 02138.}
\altaffiltext{3}{Steward Observatory, The University of Arizona, 933 N. Cherry Ave, Tucson, AZ, 85721.}

\begin{abstract}

All transiting planet  observations are at risk of contamination from nearby, unresolved stars. Blends dilute the transit signal, causing the planet to appear smaller than it really is, or producing a false positive detection when the target star is blended with an eclipsing binary. High spatial resolution adaptive optics images are an effective way of resolving most blends. Here we present visual companions and detection limits for 12 Kepler planet candidate host stars, of which 4 have companions within 4\arcsec. One system (KOI 1537) consists of two similar-magnitude stars separated by 0.1\arcsec, while KOI 174 has a companion at 0.5\arcsec. In addition, observations were made of 15 transiting planets that were previously discovered by other surveys. The only companion found within 1\arcsec\ of a known planet is the previously identified companion to WASP-2b. An additional four systems have companions between 1-4\arcsec: HAT-P-30b (3.7\arcsec, $\Delta Ks = 2.9$), HAT-P-32b (2.9\arcsec, $\Delta Ks = 3.4$), TrES-1b (2.3\arcsec, $\Delta Ks = 7.7$), and  WASP-P-33b (1.9\arcsec, $\Delta Ks = 5.5$), some of which have not been reported previously. Depending on the spatial resolution of the transit photometry for these systems, these companion stars may require a reassessment of the planetary parameters derived from transit light curves. For all systems observed, we report the limiting magnitudes beyond which additional fainter objects located 0.1-4\arcsec\ from the target may still exist. 

\end{abstract}

\keywords{binaries: general -- planets and satellites: detection -- instrumentation: adaptive optics  }

\section{Introduction}

High-resolution images are crucial to understanding the true properties of extrasolar planets. Unresolved light from additional stars near the planet host can contaminate both the transit and radial velocity signals. The result is either a false positive detection or a possibly substantial revision of the planetary parameters. An example of a significant revision in both mass and radius is Kepler-14b \citep{Buchhave2011}. The transit-hosting star has a visual companion of similar brightness separated by 0.3\arcsec, which was discovered by high-resolution imaging. Without correcting for the presence of the companion, the planetary radius would have been underestimated by 10\% and the planetary mass by 60\%.

High-resolution imaging of the Kepler planetary candidates has been done using three methods in the visible wavelength region: (a) speckle imaging \citep{Howell2011}, which has particular strength within 0.15\arcsec\ of the target star, (b) 'lucky' imaging \citep{LilloBox2012},  and (c) a current snapshot program on HST using the Wide Field Camera 3 (Program 12893, PI. R. Gilliland).  In the near infrared, a fourth method is available: adaptive optics (AO) imaging. AO images provide a useful complement because they can reveal cooler visual companions (which may have too high a contrast ratio to detect visibly), and because wider fields of view are common, can extend the detection region to further than 1\arcsec\ from the target star. (Note that we use the word "companion" to refer to \emph{visual} companions, regardless of the star(s) status as physically bound, which is unknown for the vast majority of visual companions.)

Only a few hundred of the nearly 3000 Kepler Objects of Interest (KOIs) have high resolution images in any wavelength so far, due to telescope and magnitude constraints. Recently, we published high-resolution adaptive optics images for 90 candidates discovered by the \emph{Kepler} mission \citep{Adams2012}, which were observed during the 2009-10 seasons.  Since fall 2012,  follow-up images taken by Kepler team members and the general community  have been made available on the Community Follow-On Project website\footnote{https://cfop.ipac.caltech.edu}, including the image files and tables of detection sensitivity.

In addition to the Kepler candidates, we also looked at several confirmed transiting planets discovered by other surveys. Transit surveys, by design, often use wide field of view cameras with large projected pixel sizes and even larger photometric apertures. Most surveys do not have the resources to search for nearby visual companions beyond those detected in the survey data. Follow-up high resolution images are thus important to search for blend scenarios. We note that all of the non-KOI objects in this paper have been confirmed to have planetary mass, since they have both transit photometry and radial velocity. However, additional visual companions may require a revision in the planetary parameters derived for some of these targets depending on the nature of the observations made.

In this paper, we report on 12 \emph{Kepler} candidates and 15 confirmed transiting planets which observed in 2011.

\section{Observations and data analysis}
\label{section:obs}

All observations were made in 2011 October 9-11 on the MMT, with observations of 27 unique stars with usable data. A single band (\emph{Ks}) was used to allow for more targets to be imaged. 

The Arizona Infrared imager and Echelle Spectrograph (ARIES) on the 6.5m MMT telescope can provide diffraction-limited imaging in the \emph{J} and \emph{Ks} bands \citep{McCarthy1998}. Standard Barr-MKO filters are used\footnote{http://aries.as.arizona.edu/index.php/ARIES\_Capabilities}. ARIES is fed by the adaptive secondary AO system. All objects were imaged in the f/30 mode, with a field of view of 20\arcsec\ $\times$ 20\arcsec\ and a plate scale of 0.02085\arcsec\ per pixel. The adaptive optics system in all cases guided on the primary target. Due to poor weather and some instrument problems, the image quality was highly variable. The full-width at half-maximum of the point spread functions (PSFs) was $0.23\arcsec$ at \emph{Ks}, with a best resolution of  0.1\arcsec.

For each object, one set of 16 images on a 4-point dither pattern were observed in \emph{Ks}.  Integration times varied from 0.9 to 30 s depending on stellar magnitude; in some cases, an additional set of 16 images were taken, particular when the AO lock was lost during high winds or bad seeing. Two objects taken near airmass of 2 (KOI 979 and WASP-2b) never achieved satisfactory AO lock, and have markedly elongated PSFs. However, these objects are included because they can still be searched for brighter, more distant companions. 

The images for each filter were calibrated using standard IRAF procedures\footnote{http://iraf.noao.edu/}, and combined and sky-subtracted using the \emph{xmosaic} function in the \emph{xdimsum} package. The orientations of the fields, and the absolute direction of any companions from the target star, are estimated from the dither pattern (a 2\arcsec\ square), and are only accurate to within a few degrees.

The ARIES observations only determine relative magnitudes between two stars, and does not attempt to internally calibrate to absolute magnitudes. Instead, the 2MASS catalog magnitudes (where available) are assumed to be correct, and are used to assign apparent magnitudes to fainter stars (e.g. an object with $\Delta Ks$=5 near a star with 2MASS $Ks=12$ is assigned a magnitude of $Ks = 17$). For cases where two very close stars are detected, it was assumed that the 2MASS magnitude is itself the blend of the light from both stars. The $Ks$ apparent magnitude for each star is obtained using the relative magnitudes found by ARIES, so that the sum of all apparent magnitudes is equal to the 2MASS magnitude.

The apparent magnitude of each star in $Ks$ is then converted to the Kepler magnitudes,  $Kp$, using the polynomial fits for
dwarf stars from \citet{Howell2012}.

\begin{equation}
Kp-Ks =\begin{cases}
\begin{split}
-643.05169+246.00603\emph{Ks}\\
-37.136501\emph{Ks}^2 +2.7802622\emph{Ks}^3\\
 - 0.10349091{Ks}^4 +0.0015364343\emph{Ks}^5,
\end{split}
& Ks \le 15.4, \\
-2.7284 + 0.3311\emph{Ks}, 			& Ks > 15.4.
\end{cases}
\label{eqnKs}
\end{equation}

The single-filter conversion in Equation~\ref{eqnKs} yields $Kp$ magnitude estimates that are accurate to approximately $0.6$-$0.8$ mag, and are used to estimate the magnitude of detected companions in the $Kepler$ wavelength band.

\section{Detection limits}
\label{section:limits}

Companion objects were identified individually by visual inspection. The processed images contain numerous artifacts, including instrumental artifacts near the edges and in a box pattern 1/2 of the CCD size away from bright stars, as well as speckles near the PSF. With a little experience, the artifacts are easy to recognize and reject visually. The speckle pattern changes slowly over the night, and images of other targets are used to distinguish the true stars from noise. Once companions are identified, the magnitude of each star is found using the IRAF routine \emph{phot} with a 5 pixel aperture (large enough to capture most of the PSF without including light from nearby companions). For stars with companions within 0.5\arcsec, the relative magnitudes are estimated by PSF fitting all stars in a portion of the image using routines written in Mathematica.

Upper limits on detectable stars are taken as 5 times the standard deviation above the background level. This level is determined for various distances from the target star by constructing a series of concentric annuli centered on the target star. A separate calculation using the IRAF routine \emph{phot} is used to determine the mean counts and standard deviation within each annulus using the settings in \emph{fitskypars}; the mean counts at distance (10\arcsec) are subtracted from the background counts for each annulus so that only the counts from the star remain, and the standard deviation is used to calculate the 5$\sigma$ limits. The widths of the annuli are set so that they do not overlap: 0.05\arcsec\ wide for annuli up to 0.2\arcsec\ from the star; 0.1\arcsec\ wide between 0.2-1.0\arcsec, and 1\arcsec\ wide for annuli from 1\arcsec\ outward. By 4\arcsec\ the maximum limiting magnitude has been reached, and more distant objects were also detected, though with incomplete spatial coverage beyond 5-10\arcsec\ depending on where the star falls relative to the edges of the chip. Limits are reported for annuli at distances of  0.1-4\arcsec\ in Table~\ref{table:limits}.

The innermost detectable object is a function of the observed PSF of the target star. The best FWHM achieved was 0.1\arcsec\ in \emph{Ks}. However, poor weather and problems with the AO systems often caused excursions well above that level. The magnitude limits for each object are shown in Table~\ref{table:limits}. 

\section{Visual companions to Kepler candidates}

\begin{figure}
\includegraphics*[scale=1]{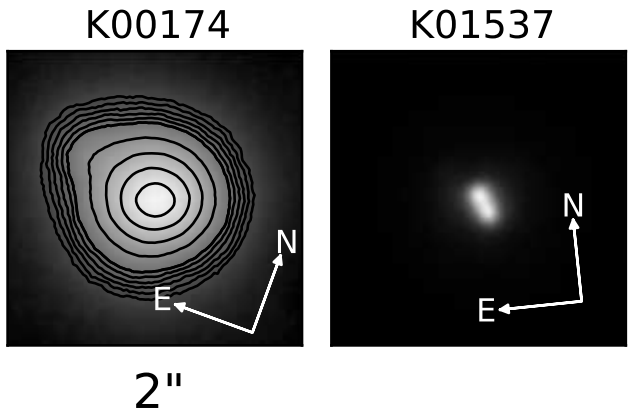}
\caption{Adaptive optics images of targets with companions closer than 0.5\arcsec. Each field of view is 2\arcsec\ on a side. The faint, blended companion to KOI 174 is highlighted with contours.}
\label{fig:near}
\end{figure}

\begin{figure}
\includegraphics*[scale=0.8]{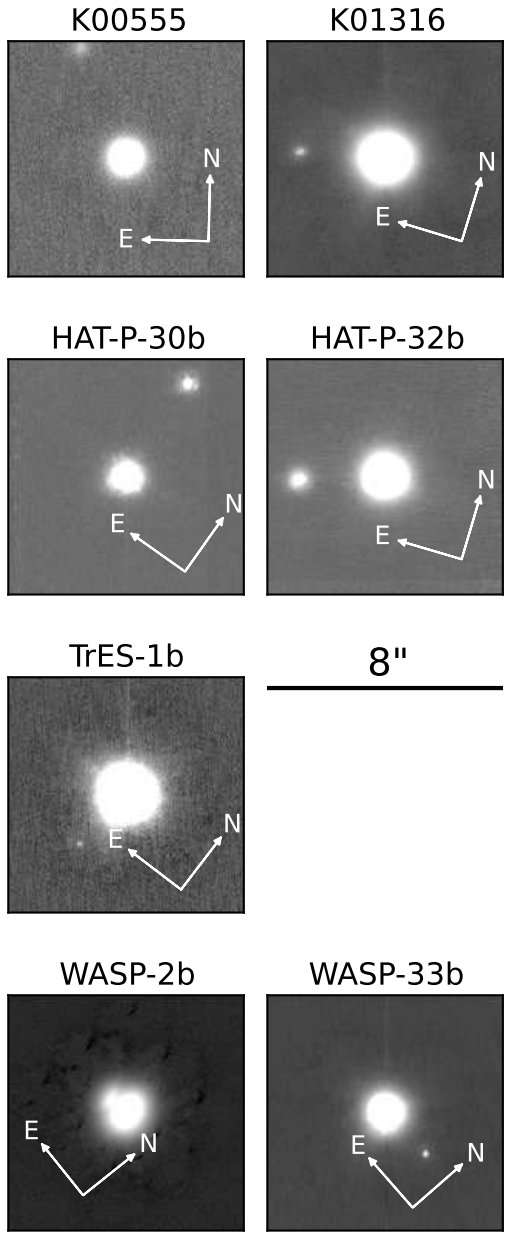}
\caption{Adaptive optics images of targets with companions from 0.5-4\arcsec. Each field of view is 8\arcsec\ on a side. }
\label{fig:far}
\end{figure}

All visual companion stars detected within 0.5\arcsec\ of the primary target are shown in Figure~\ref{fig:near} and within 0.5-4\arcsec\ in Figure~\ref{fig:far}. Four Kepler candidates with such companions are discussed below, and five companions to confirmed planets are discussed in the next section. The number of companions found within 2\arcsec\ is 2 out of 12, or 17\%, consistent with the value of 20\% found in \citet{Adams2012}.

\subsection{KOI 174}

KOI 174 has a $1.94~R_E$ candidate planet \citep{Batalha2012}.  The stellar companion detected with AO in this work is only 0.5\arcsec\  away, with $\Delta Ks = 2.6$. We note that no companions were detected in visible wavelength speckle data \citep{Howell2011}, which can typically detect objects 3-4 mag fainter. The estimated magnitude difference in visible wavelengths, $\Delta Kp = 2.3$, is nominally within the Speckle detection range, but given the somewhat different filters used and the uncertainties in magnitude transformation using a single band it is not unreasonable that this object  was not seen.

No significant centroid shift during transit relative to the out-of-transit position is seen in the Q1-Q12 data validation report\footnote{DV reports for all KOIs in this paper are available for download at http://exoplanetarchive.ipac.caltech.edu/}, indicating that the companion is around the target star and unlikely to be causing a false positive detection \citep{Bryson2013}. The centroid location does differ from the Kepler Input Catalog (KIC) position by 3.9$\sigma$, or 0.6\arcsec, which is  not surprising given the unresolved nearby star.

However, the planetary parameters need to be revised to account for the extra light in the Kepler wavelength. A dilution correction of about 10\% should be applied to the transit radius, depending on the magnitude of the companion in the Kepler wavelength. If radial velocity measurements are obtained of this system ($Kp=13.779$), they will also need to be corrected for the companion star's contribution. Additional, deeper high-resolution images in other wavelengths are encouraged to resolve the status of this system.

\subsection{KOI 555}

KOI 555 hosts two planet candidates, with radii of $1.41~R_E$ and $2.68~R_E$ \citep{Batalha2012}.  A  stellar companion at 4\arcsec\ was detected using lucky imaging, and was found to have $\Delta i=3.58$ \citep{LilloBox2012}. We find a similar magnitude difference, with $\Delta Ks = 3.43$. This object was detected in UKIRT images with $J=16.765$ ($\Delta J = 3.36$) and is listed as a probable galaxy in the source table on the CFOP website. This faint, distant companion contributes at most a few percent dilution to transit observations, depending on the photometric aperture used. 

No additional, closer targets were detected, and no significant centroid shift relative to out-of-transit is seen in the Q1-Q12 data validation report, ruling out a false positive interpretation.

\subsection{KOI 1316}

KOI 1316 has two planet candidates \citep{Borucki2012, Batalha2012}, both $1.47~R_E$. Our AO images reveal a pair of faint companion stars, both with $\Delta \text{Ks}=6$, located 2.9\arcsec\ from the primary target and about 0.16\arcsec\ from each other. Due to the small size of the planets around KOI 1316, the Kepler pipeline only detected a single planet automatically, with low signal-to-noise and a modest $3~\sigma$ centroid shift during transit ($0.8 \pm 0.215$\arcsec). However, the offset may be due to measurement or crowding bias in the location of the target star. Also, the direction and magnitude of the shift (about 0.5\arcsec\ S and 0.5\arcsec\ E) do not precisely correspond to the location of the two faint companion stars (roughly 3\arcsec\ E). A more thorough modeling of the centroid of this system that accounts for all observations is recommended. Both stars together are estimated to contribute much less than 1\% of the starlight in the Kepler aperture, implying only a modest dilution correction if the system is not a false positive.

\subsection{KOI 1537}

KOI 1537 has another small planetary candidate with $R_p=0.85~R_E$ according to \citet{Batalha2012}. However, this value should be revised upward in light of the close companion revealed in the AO images. In \emph{Ks} band the two stars are nearly identical in brightness, with the fainter ($\Delta Ks = 0.15$) companion located 0.13\arcsec\ and at an angle (N to E) of $159.5^{\circ}$ from the primary. The estimated Kepler magnitude of this star is $Kp =12.9$, although for single color transformations the errors are large ($0.6-0.8$ mag). In the worst case scenario where the magnitudes were the same in the Kepler bandpass, the true transit radius is larger by about a factor of 1.4, making the planet radius $R_p\sim1.2 R_E$. 

We note that this star is unclassified in the Kepler Input Catalog, and the planetary radius in \citet{Batalha2012} was calculated assuming it is on the main sequence and has the effective temperature defined by the J-K color. The nearby visual companion may have contributed to difficulties in characterizing this target. A self-consistent reanalysis of the stellar and planetary parameters is recommended using the AO and other follow up observations.

\section{Companions to stars with known transiting planets}

\subsection{HAT-P-30b}

HAT-P-30b was independently discovered by both the HAT \citep{Johnson2011} and WASP surveys \citep[aka WASP-51,][]{Enoch2011}. The companion detected with AO in this work is relatively distant, 3.7\arcsec\ with $\Delta Ks = 2.9$. We found no evidence of the "faint" companion that \citet{Enoch2011} observed at 1.5\arcsec\ with the guide star camera. At that distance, our data would have been sensitive to companions as faint as $\Delta Ks = 9.2$, implying that this companion must be quite faint in near-infrared wavelengths. The companion we did detect is outside of the 3\arcsec\ fiber width used by \citet{Enoch2011} and may lie outside the typical aperture used for photometric observations.

\subsection{HAT-P-32b}

HAT-P-32b was discovered by \citet{Hartman2011b}, and is a short-period hot Jupiter with a highly inflated radius. The discovery paper notes that images taken in $i$ band rule out companions with $\Delta i \le 5$ and separations greater than 3\arcsec, or $\Delta i \le 2$ and separations greater than 1\arcsec. The companion detected with AO in this work falls near the edges of those limits, with a distance of 2.9\arcsec\ and a magnitude difference of $\Delta Ks = 3.4$. This is consistent with the visible constraints if the companion is a bit fainter in $i$ than in \emph{Ks}.  Based on the analysis done by \citet{Hartman2011b} to rule out false positive scenarios, the most likely interpretation is a faint companion that contributes a modest amount of dilution to the overall planetary system. A star with $\Delta Ks = 3.4$ contributes about 4\% of the light in \emph{Ks} (this work) and no more than about 1\% in $V$ using the observational constraints that $\Delta i > 5$ from \citet{Hartman2011b}. Thus, the value for the planetary radius derived from transit photometry should be adjusted upward modestly, making the planet slightly more inflated than initially reported.

\subsection{TrES-1b}

TrES-1b was one of the first transiting planets discovered \citep{Alonso2004}, and has been observed in many wavelengths on numerous ground and space based platforms. We achieved excellent limits on this object ($\Delta Ks = 3$ mag at 0.2\arcsec\ to $\Delta Ks = 10$ mag at 4\arcsec). The sole companion detected by AO imaging is located 2.3\arcsec\ away and has $\Delta Ks = 7.7$. Multi-wavelength images would be needed to estimate the brightness of this object in other observing bands, and hence to estimate the possibility of dilution in existing observations. We note that even objects as faint as this one can affect very high-precision observations: if fully blended, this object would contribution 800 ppm to the transit depth in \emph{Ks}, potentially detectable in high-precision, wide-aperture observations intended, for instance, to detect slight changes in the transit radius at different wavelengths due to the planetary atmosphere.

\subsection{WASP-2b}
WASP-2b was discovered by \citet{CollierCameron2007}. In the discovery paper, AO images were reported using the NAOMI adaptive-optics system on the 4.2-m William Herschel Telescope. They found a single companion with $\Delta H = 2.7$ mag at 0.7\arcsec, with corrected FWHM of 0.2\arcsec. Later observations by \citet{Daemgen2009} using Lucky Imaging found finding values of $\Delta i'=4.095$ and $\Delta z' =3.626$, with a distance of 0.757\arcsec\ and PA=104.7. This companion was also detected in our images despite poor AO conditions (note the elongated shapes, caused by observing at high airmass, in Figure~\ref{fig:far}). Our AO images provide the first \emph{Ks} detection of the companion at 0.7\arcsec, with $\Delta Ks = 2.3$. No additional companions were found, to the limits provided in Table~\ref{table:limits}.

\subsection{WASP-33b}

WASP-33b, one of the hottest-known exoplanets \citep{Christian2006, CollierCameron2010}, was reported  by \citet{Moya2011} to have a companion shortly after these observations were made. The parameters reported in that paper, 1.961\arcsec\ with $\Delta Ks = 5.69$, agree with the values found in this work.

\section{Conclusions}
\label{section:ogle56conclusions}

We have observed 12 Kepler candidates and 15 previously known transiting planets. Companions were discovered within 1\arcsec\ for two of the Kepler targets, KOI 174 and KOI 1537, and observed for the first time in \emph{Ks} for WASP-2b. Additional companions were found or confirmed between 1-4\arcsec\ for the following stars: KOI 555, KOI 1316, HAT-P-30b, HAT-P-32b, TrES-1b, and WASP-33b. The number of companions found within 2\arcsec\ for Kepler targets, 2 out of 12 or 17\%, is consistent with the value of 20\% found in \citet{Adams2012}. As noted in that paper, many of the closest companions (within 2\arcsec) are expected to be physically bound since their frequency is uncorrelated with Galactic latitude, but further work is needed to determine the status of an individual system.

The impact of these companion stars depends on the object separation and the resolution of the observations. For the companions at 2-4\arcsec, all detected companions are relatively faint, and the expected effect is a potential minor dilution affecting the derived transit depths for KOI 555, HAT-P-30b, HAT-P-32b, TrES-1b, and WASP-33b (depending on the aperture and resolution of the observations in question). 

Objects with close companions will require more extensive corrections. KOI 174 will need a dilution correction of about 10\%, depending on the colors of the blending object in visible wavelengths.  KOI 1316 needs additional examination of the centroid shifts to determine if it is a false positive. The presence of another, nearly identical star within 0.12\arcsec\ of KOI 1537 does not invalidate the planetary nature of its candidate, but does mean it may be as much as 1.4 times larger than the reported transit radius (i.e., $1.2~ R_E$ instead of $0.85~ R_E$).

No companions were detected closer than 4\arcsec\ around the following stars, with inner detection limits in parentheses:  KOIs 341 (0.5\arcsec), 638 (0.2\arcsec), 700 (1\arcsec), 961 (0.5\arcsec), 973 (0.2\arcsec), 979 (1\arcsec), 1054 (0.5\arcsec), and 1883 (0.5\arcsec), and confirmed transiting planets Corot-1b (0.5\arcsec),  HAT-P-17b (0.5\arcsec),  HAT-P-25b (0.5\arcsec),  HAT-P-33b (0.2\arcsec),  HAT-P-6b (0.2\arcsec),  HAT-P-9b (0.2\arcsec), HD17156b (0.5\arcsec), WASP-1b (0.5\arcsec), XO-3b (0.1\arcsec), and XO-4b (0.2\arcsec).

All images and limit tables for the Kepler candidates presented in this paper are available on the Community Follow-On Project website (https://cfop.ipac.caltech.edu/).

\acknowledgements

Authors gratefully acknowledge partial support from NASA grant NNX10AK54A. Thanks also go to David Ciardi for coordination of the Kepler Follow-on Program (KFOP), and to Steve Bryson for helpful conversations about centroid offsets. We thank an anonymous referee for his/her suggested improvements.

{\it Facilities:}  \facility{MMT (ARIES)}, \facility{Kepler}


\bibliographystyle{apj}
\bibliography{main}


\newpage

\begin{deluxetable}{c c c c   c c c c c c }
\tablecolumns{11} 
\tablewidth{0pt}
\tabletypesize{\scriptsize}
\tablecaption{Limits on nearby stars for all objects}
\tablehead{ 
KOI  	&KeplerID 	& FWHM & Ks (2MASS)		&  \multicolumn{6}{c}{Limiting $\Delta$mag for annulus centered at ...} \\ 
	&			& (\arcsec)	&		& 0.1\arcsec& 0.2\arcsec\ &  0.5\arcsec\ &  1\arcsec\ & 2\arcsec\ &  4\arcsec\  
}
\startdata
KOI 174		 & 10810838	 &0.42	 &11.536 	& --  	& --  	& 1.47 	& 4.0 	& 4.27 	& 5.7 \\
KOI 341		 & 10878263	 &0.35	 &11.698 	& --  	& --  	& 1.89 	& 4.43 	& 4.84 	& 6.22 \\
KOI 555		 & 5709725	 &0.44	 &12.931 	& --  	& --  	& 1.36 	& 3.63 	& 3.82 	& 4.24 \\
KOI 638		 & 5113822	 &0.18	 &12.083 	& --  	& 1.61 	& 3.02 	& 5.43 	& 5.79 	& 6.55 \\
KOI 700		 & 8962094	 &0.53	 &11.996 	& --  	& --  	& --  	& 3.44 	& 3.69 	& 4.68 \\
KOI 961		 & 8561063	 &0.4	 &11.465 	& --  	& --  	& 1.66 	& 4.45 	& 4.86 	& 6.25 \\
KOI 973		 & 12366681	 &0.16	 &8.366 	& --  	& 1.95 	& 3.38 	& 6.16 	& 6.58 	& 8.83 \\
KOI 979		 & 5648449	 &0.53	 &6.280 	& --  	& --  	& --  	& 3.37 	& 3.83 	& 7.21 \\
KOI 1054		 & 6032981	 &0.24	 &10.066 	& --  	& --  	& 2.96 	& 5.7 	& 6.09 	& 7.68 \\
KOI 1316		 & 10794087	 &0.16	 &10.562 	& --  	& 1.77 	& 3.13 	& 5.65 	& 6.04 	& 8.44 \\
KOI 1537		 & 9872292	 &0.21	 &10.514 	& --  	& --  	& 3.43 	& 6.18 	& 6.57 	& 7.59 \\
KOI 1883		 & 11758544	 &0.22	 &10.608 	& --  	& --  	& 2.21 	& 4.45 	& 4.75 	& 7.07 \\
Corot-1b		 & --	 		&0.35	 &12.149 	& --  	& --  	& 6.14 	& 7.05 	& 7.13 	& 7.01 \\
HAT-P-17b	 & --	 		&0.23	 &8.544 	& --  	& --  	& 2.85 	& 5.1 	& 7.77 	& 9.23 \\
HAT-P-25b	 & --	 		&0.49	 &10.815 	& --  	& --  	& 1.88 	& 4.27 	& 6.79 	& 7.78 \\
HAT-P-30b	 & --	 		&0.11	 &9.151 	& --  	& 3.74 	& 5.75 	& 8.32 	& 9.56 	& 9.5 \\
HAT-P-32b	 & --	 		&0.12	 &15.419 	& --  	& 2.68 	& 4.35 	& 6.75 	& 8.94 	& 9.46 \\
HAT-P-33b	 & --			 &0.12	 &10.004 	& --  	& 2.75 	& 4.11 	& 6.3 	& 8.75 	& 9.22 \\
HAT-P-6b		 & -- 			&0.15	 &9.313 	& --  	& 2.24 	& 3.83 	& 6.25 	& 8.97 	& 10.44 \\
HAT-P-9b		 & -- 			&0.14	 &11.015 	& --  	& 2.55 	& 4.25 	& 6.81 	& 8.37 	& 8.46 \\
HD17156b	 & --			&0.35	 &6.763 	& --  	& --  	& 3.66 	& 6.04 	& 8.74 	& 10.14 \\
TrES-1b	 	 & --			&0.13	 &15.611 	& --  	& 2.8 	& 4.55 	& 7.06 	& 9.21 	& 9.99 \\
WASP-1b	 	 & --			&0.28	 &10.276 	& --  	& --  	& 2.85 	& 5.32 	& 8.03 	& 9.23 \\
WASP-2b	 	 & --			&0.46	 &9.632 	& --  	& --  	& 2.05 	& 4.22 	& 7.19 	& 7.93 \\
WASP-33b	 & --			&0.23	 &7.468 	& --  	& --  	& 4.68 	& 7.18 	& 9.45 	& 10.13 \\
XO-3b		 & --			 &0.1	 &8.791 	& 2.25 	& 3.23 	& 4.87 	& 7.24 	& 9.61 	& 10.11 \\
XO-4b		 & -- 			&0.11	 &9.406 	& --  	& 3.04 	& 4.37 	& 6.76 	& 8.66 	& 9.04 \\
\enddata     
\label{table:limits}
\end{deluxetable}

\newpage

\begin{deluxetable}{rrrrrr    rrr    }
\tablewidth{0pt}
\tabletypesize{\scriptsize}
\tablecaption{Additional stars within 4\arcsec\ of target}
\tablehead{
KOI	&KeplerID	 &Kp	& 2MASS (K)	& Star\tablenotemark{a} &Dist(\arcsec)	&PA($^{\circ}$)\tablenotemark{b}	&$\Delta$Mag ($Ks$)\tablenotemark{c}  & Mag (\emph{Kp})\tablenotemark{d}
		}
\startdata
K00174		&10810838	&13.779	&11.536	&1	&0.51\tablenotemark{e}	&70.2\tablenotemark{e}	&2.57\tablenotemark{e}	&16.1\tablenotemark{e} \\
K00555		&5709725	&14.759	&12.931	&1	&4.01	&20		&3.43	&19.0 \\
K01316		&10794087	&11.926	&10.562	&1	&2.85	&102	&5.99	&19.3 \\
			&			&		&		&2	&2.89	&103	&5.97	&19.3 \\
K01537		&9872292	&11.740	&10.514	&1	&0.13\tablenotemark{e}	&160\tablenotemark{e}	&0.15\tablenotemark{e}	&13.0\tablenotemark{e} \\
\hline
HAT-P-30b	&--			&--		&9.151	&1	&3.74	&2		&2.92	&-- \\
HAT-P-32b	&--			&--		&15.419	&1	&2.87	&109	&3.38	&-- \\
TrES-1b		&--			&--		&15.611	&1	&2.31	&174	&7.68	&-- \\
WASP-2b		&--			&--		&9.632	&1	&0.69	&99		&2.3		&-- \\
WASP-33b	&--			&--		&7.468	&1	&1.91	&268	&5.45	&-- 	
\enddata  
\tablenotetext{a} {Each unique companion star detected is numbered. Note that for K01316, there are two nearly-identical brightness objects separated from each other by approximately the FWHM (0.16\arcsec) of the image.}
\tablenotetext{b} {Angle from north, determined from the north-east directions of the 2\arcsec\ square dither pattern; angles may differ from true N-E by a few degrees.}
\tablenotetext{c} {Error on $\Delta$Mag is about 0.01 mag.}
\tablenotetext{d} {$Kp$ magnitude estimated for a dwarf companion using Equation~\ref{eqnKs}.}
\tablenotetext{e} {Offset and delta magnitude for objects with separations close to the FWHM were found using PSF fitting.}
   
\label{table:compstars}
\end{deluxetable}

\clearpage

\end{document}